 \documentclass[final,5p,times]{elsarticle}

\usepackage{amssymb}
\usepackage{amsmath,amssymb,amsfonts}
\usepackage{algorithmic}

\usepackage{threeparttable}
\usepackage{algorithm,algorithmic}
 \usepackage{multirow}
 \usepackage{float}
 \usepackage{booktabs}
\usepackage[font=small,labelfont=bf,labelsep=none]{caption}  
\begin{document}
	
	\begin{frontmatter}
		
		
		
		\title{Automatic Ultrasound Curve Angle Measurement via Affinity Clustering for Adolescent Idiopathic Scoliosis Evaluation}
		
		
		\author[]{Yihao Zhou$^{a}$}

		\author[]{Timothy Tin-Yan Lee$^{a}$}
		\author[]{Kelly Ka-Lee Lai$^{a}$}
		\author[]{Chonglin Wu$^{a}$}
		\author[]{Hin Ting Lau$^{a}$}
		\author[]{De Yang$^{a}$}
		\author[]{Chui-Yi Chan$^{a}$}
		\author[]{Winnie Chiu-Wing Chu$^{d}$}
		\author[]{Jack Chun-Yiu Cheng$^{b,c}$}
		\author[]{Tsz-Ping Lam$^{b,c}$}
		\author[]{Yong-Ping Zheng$^{a}$}
		
		\affiliation[1]{Department of Biomedical Engineering, The Hong Kong Polytechnic University, Hong Kong, China}
		\affiliation[2]{Department of Orthopaedics and Traumatology, The Chinese University of Hong Kong, Hong Kong SAR, China}
		\affiliation[3]{SH Ho Scoliosis Research Lab, Joint Scoliosis Research Center of the Chinese University of Hong Kong and Nanjing University}
		\affiliation[4]{Department of Imaging and Interventional Radiology, The Chinese University of Hong Kong, Hong Kong, China}

		\begin{abstract}
			The current clinical gold standard for evaluating adolescent idiopathic scoliosis (AIS) is X-ray radiography, using Cobb angle measurement. However, the frequent monitoring of the AIS progression using X-rays poses a challenge due to the cumulative radiation exposure. Although 3D ultrasound has been validated as a reliable and radiation-free alternative for scoliosis assessment, the process of measuring spinal curvature is still carried out manually. Consequently, there is a considerable demand for a fully automatic system that can locate bony landmarks and perform angle measurements. To this end, we introduce an estimation model for automatic ultrasound curve angle (UCA) measurement. The model employs a dual-branch network to detect candidate landmarks and perform vertebra segmentation on ultrasound coronal images. An affinity clustering strategy is utilized within the vertebral segmentation area to illustrate the affinity relationship between candidate landmarks. Subsequently, we can efficiently perform line delineation from a clustered affinity map for UCA measurement. As our method is specifically designed for UCA calculation, this method outperforms other state-of-the-art methods for landmark and line detection tasks. The high correlation between the automatic UCA and Cobb angle (R$^2$=0.858) suggests that our proposed method can potentially replace manual UCA measurement in ultrasound scoliosis assessment. 
			
		\end{abstract}
		
		
		
		
	\end{frontmatter}
	
	
	\section{Introduction}
	\label{}
	Adolescent idiopathic scoliosis (AIS) is the most prevalent spinal deformity in children, affecting approximately 0.47-5.2$\%$ of teenagers \cite{Konieczny2013}. Clinical diagnosis and progression monitoring of scoliosis rely on the radiographic Cobb measurement of the spine. However, frequent use of X-rays is not viable due to the potentially harm from cumulative radiation exposure, especially in patients who require regular curve monitoring \cite{Simony2016, Himmetoglu2015, McArthur2015}.
	Though EOS imaging system offers low-dose radiographs, its high setup cost limits its widespread adoption \cite{jeon2018combination}. In addition, it is impractical for resource-constrained healthcare facilities to utilize such system. Therefore, it is essential to explore alternative imaging modalities that are cost-effective, safe, and easily accessible for more regular scoliosis monitoring.
	\begin{figure}[t]
		\centering
		\includegraphics[height=0.22\textheight, width=0.45\textwidth]{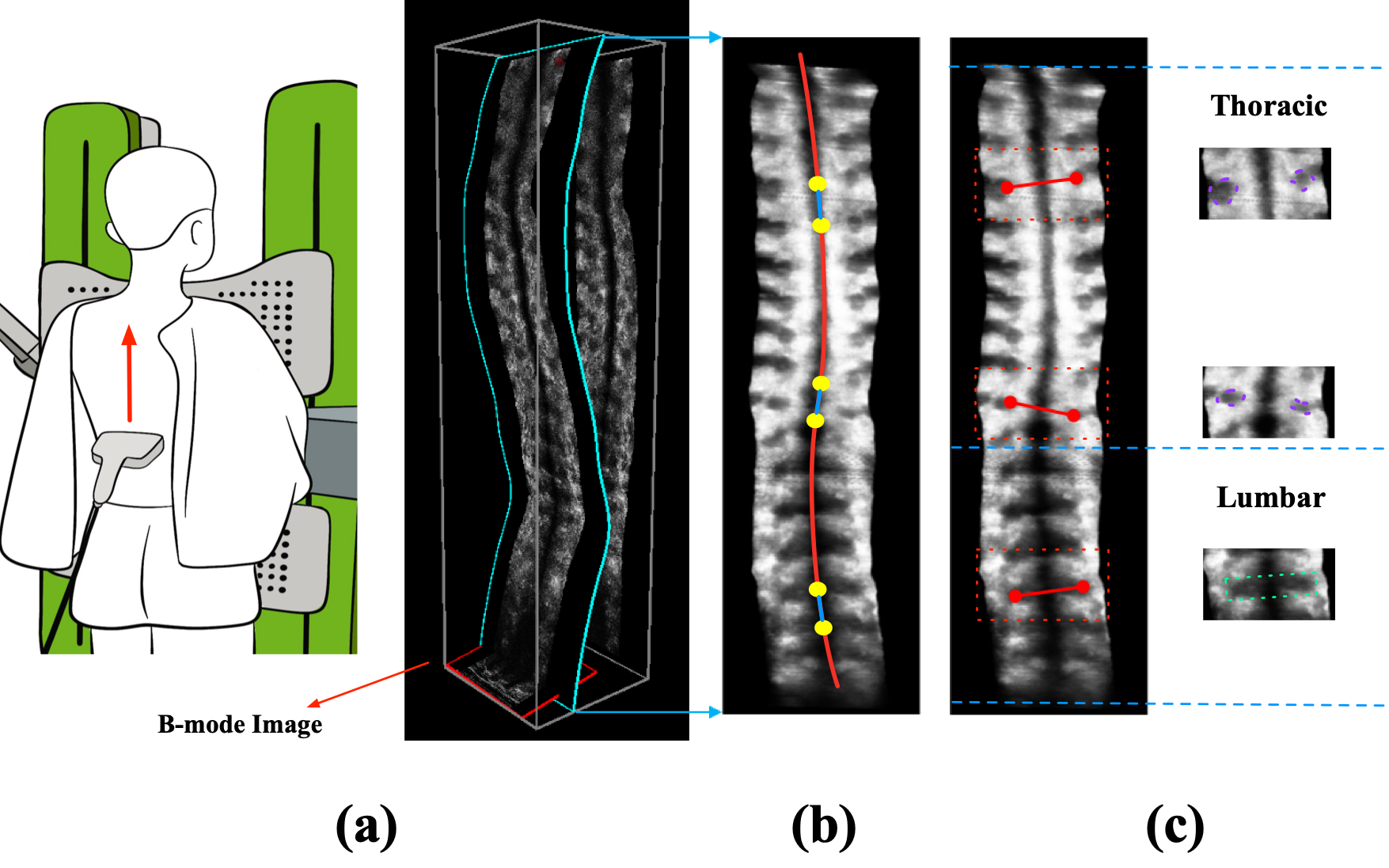}
		\caption{. (a) An Illustration of the generation of a volume projection imaging (VPI). The probe is being moved at a constant speed from bottom to top along the spine curve on the patients' skin. B-mode images combined with recorded spatial information are grouped for the generation of 3D ultrasound volume. The coronal ultrasound image is then generated using the VPI method, which incorporates a customized depth profile based on the distance from the skin to the laminae. (b) Ultrasound spinous process angle (SPA). The scoliotic curve on the medial shadow of the spinous processes is used to measure the angle for AIS diagnosis. \cite{Cheung2015_VPI} (c) Ultrasound curve angle (UCA). For thoracic region, line is placed on the center of the shadow of a transverse process (purple dotted line). For lumbar region, lines are drawn towards the center of the bilateral sides of lump (green dotted line) \cite{Lee2021}.
		}
		\label{introduction_scolioscan}
	\end{figure}	
		\begin{figure}[t]
		\centering
		\includegraphics[height=0.22\textheight, width=0.46\textwidth]{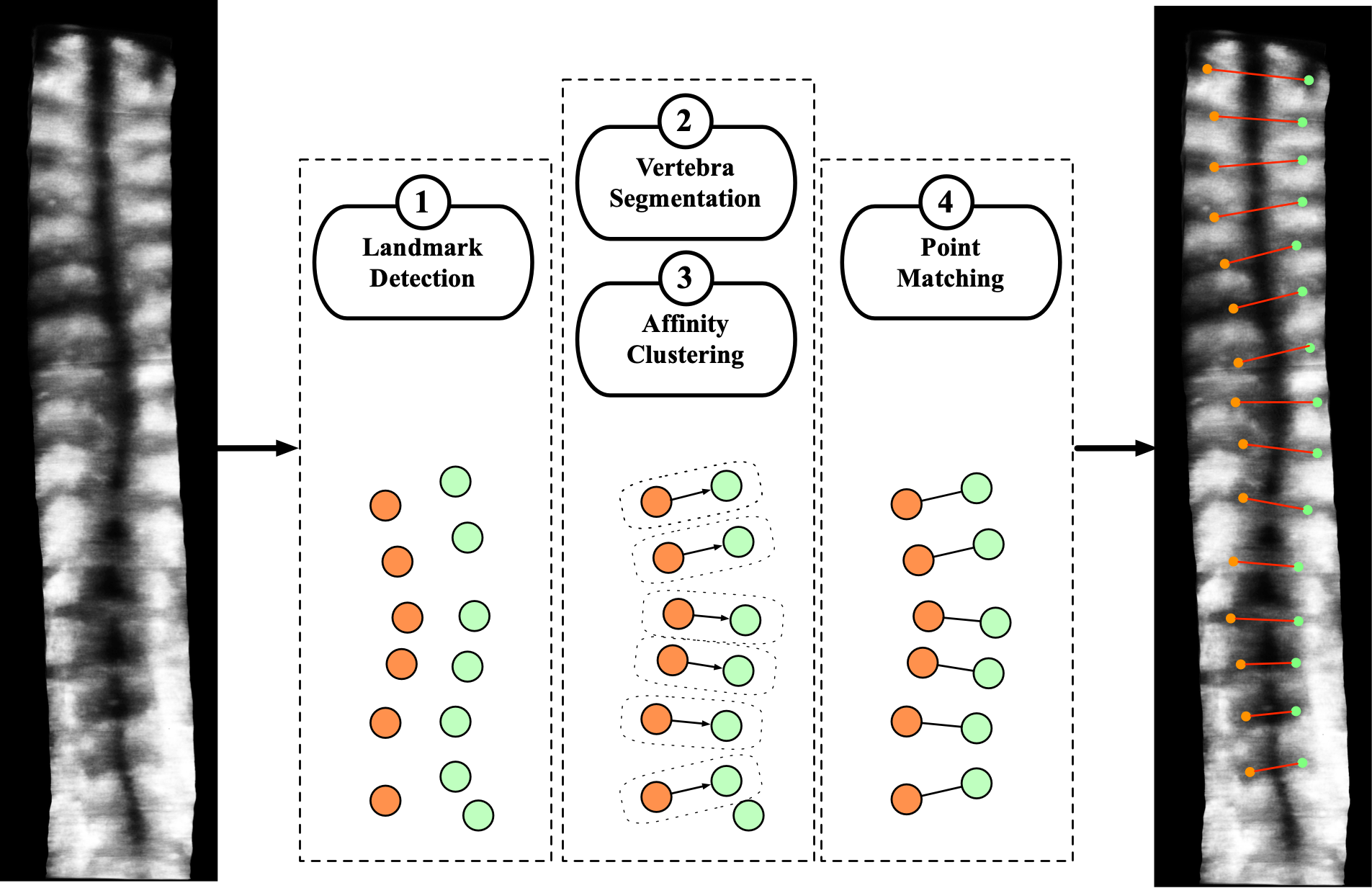}
		\caption{. The model identifies all potential anatomical landmarks along both sides of the bone curvature. The landmarks corresponding to the same vertebrae are connected based on the clustered affinity map. The most tilted lines in different regions are selected to form the UCA for the assessment of scoliosis.}
		\label{introduction_uca_pipeline}
	\end{figure}

Recently, 3D ultrasound imaging has emerged as a promising complementary imaging modality for tracking scoliosis, providing a radiation-free solution to reveal pathology. As shown in Figure.1, the subject is being scanned using an ultrasound probe, to capture a series of B-mode ultrasound images along with their corresponding 3D spatial information, thereby forming volume data. The volume projection imaging (VPI) is employed to generate 2D coronal-plane images from the volume data through non-planar volume rendering \cite{Cheung2015, Cheung2015_VPI}. The shadow of the superficial bone surface enables clinicians to observe spinal deformity due to the nature of ultrasound imaging.
	Chen et al. were the first to evaluate scoliosis in VPI images by manually identifying the spinous column profile (Figure.1.(b))  \cite{Chen2013}. Zhou et al. achieved automatic spinous curvature evaluation by utilizing prior knowledge of vertebral anatomical structures \cite{Zhou2020}. However, for patients with severe scoliosis, their spinal processes may deform and rotate significantly. Thus, the spinal profile formed by the spinous process cannot accurately represent the actual lateral deformity of the spine, leading to underestimation of spinal deformity. To evaluate the spine deformity more accurately via VPI, the ultrasound curve angle (UCA) has been proposed (Figure.1 (c)) \cite{Lee2021}. Shadows corresponding to the transverse processes and ribs in the thoracic region, as well as those from the superior and inferior articular processes in the (thoraco)lumbar region, can be identified in a similar manner. Precise recognition of these structures' foundations is crucial for line delineation. However, this manual identification process heavily relies on clinicians' experience, introducing subjectivity into the measurement. Moreover, in circumstances where image quality is severely compromised, the manual placement of lines underscores the need for trained medical professionals to interpret these images, revealing inherent limitations and potential inaccuracies in estimation. These observations motivate the development of an efficient and automated approach to identifying landmarks, which facilitates line placement and subsequent angle measurement. Accurately locating the pertinent landmarks of the vertebra is paramount for UCA measurement. We observe the pattern where landmarks manifest in pairs on both sides of the spinous profile, prompting us to approach the task like pose estimation i.e., connecting corresponding joint points \cite{Dang2019, Wang2021_survey}. As illustrated in Figure.2, our proposed method includes the following stpng:
	\begin{figure*}
		\centering
		\includegraphics[width=0.98\textwidth]{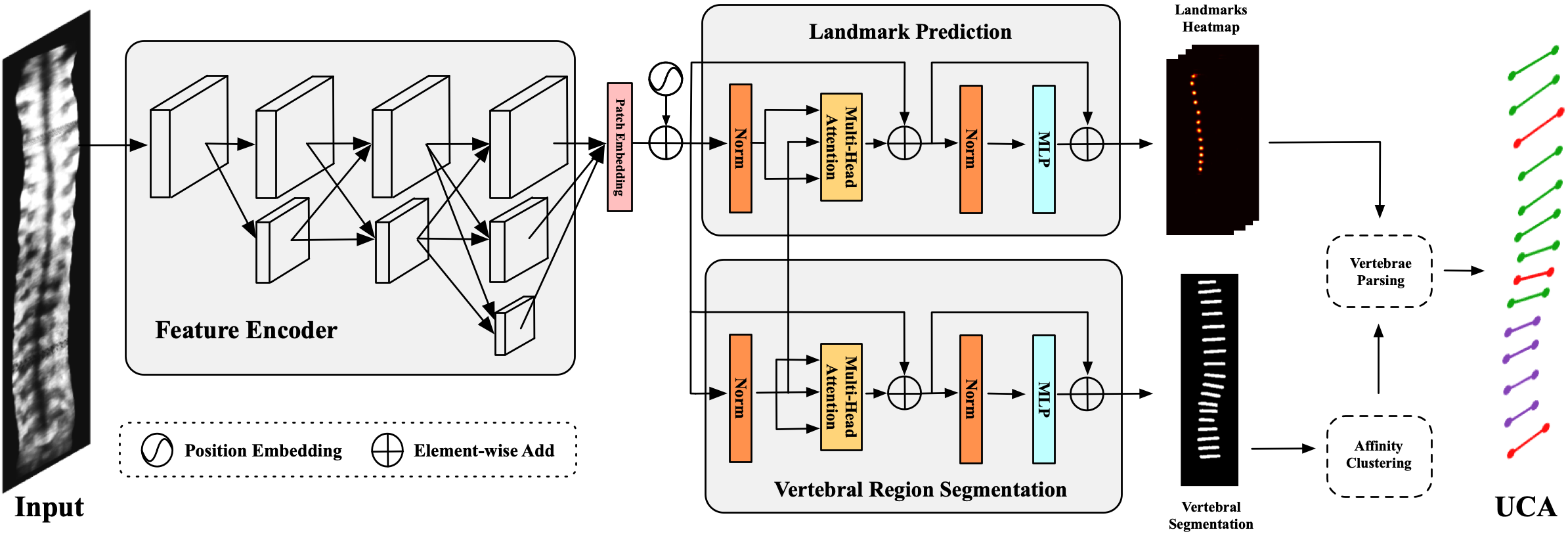}
		\caption{. Overview pipeline of automatic UCA measurement. The model extracts latent features through a feature backbone for both landmark detection and spine segmentation, respectively. The features in the segmentation decoder are shared with the landmark detection decoder. In the reference stage, the vertebrae segmentation map is parsed to represent the affinity relationship among the detected candidate points. Points belonging to the same vertebra are grouped together to form lines.}
		\label{Fig3}
	\end{figure*}
	\begin{itemize}
		\item [1)] 
		Identify anatomical landmarks and other pertinent points of interest.    
		\item [2)]
		Establish an affinity relationship between landmarks that require a connection through lines.
		\item [3)]
		Utilize a grouping strategy to associate detected landmarks with desired connections. This involves categorizing landmarks belonging to the same vertebra and differentiating them from landmarks of other vertebrae. 
		\item [4)]
		Draw lines between connected landmarks to visualize and measure the overall spine deformity.
	\end{itemize}
	Building upon these concepts, we present an estimation model for automatic UCA measurement. The model architecture comprises a dual-branch network for landmark detection and vertebra discrimination. The detection decoder predicts the heatmap of landmark locations in both thoracic and lumbar regions. Vertebra discrimination is achieved through segmentation, followed by an affinity clustering strategy that aims for affinity regression of the vertebral region on the segmentation map to align candidate landmarks. The main contributions of this study are summarized as follows:
	\begin{itemize}
		\item We have successfully achieved the automatic measurement of UCA by assembling the points to the line directly, which is reported for the first time.
		\item We have introduced an innovative affinity clustering strategy designed to capture the affinity relationships among candidate landmarks within the vertebrae segmentation map. This approach facilitates the grouping of landmarks belonging to the same vertebra, forming the angle through optimal parsing with the clustered affinity map.
		
		\item We have conducted quantitative experiments on a dataset of ultrasound coronal images with corresponding biplanar radiographs. The strong correlations with the Cobb angle illustrate that our proposed automatic method holds the potential to replace manual UCA measurements in the ultrasound assessment of scoliosis.
	\end{itemize}

	\section{Related work}
	Several previous studies have explored the use of ultrasound in diagnosing scoliosis. Cheung et al. first reported using VPI method on a sequence of 2D B-mode ultrasound images to visualize spine anatomy \cite{Cheung2015_VPI}. The VPI-SP, midline shadow curve generated by spinous processes (SPs), has demonstrated a good correlation with the Cobb angle \cite{Zheng2016, Wong2019}. Huang et.al. developed a method for real-time tracking of SPs in the ultrasonic video to establish a 3D spinal profile for deformity assessment \cite{Huang2023}. As the spine rotates, however, the curvature of SPs might be underestimated. An alternative and more accurate method, UCA, has been demonstrated to be comparable to the conventional Cobb angle \cite{Lee2021}. It computes spinal deformity using the lateral shadow features of transverse processes (TPs), articular processes, and laminae. The prevailing method for performing UCA is through manual measurement, which relies on human discretion. Some research has been conducted on spine segmentation to achieve automatic UCA measurement. Yang et.al. proposed a semi-automatic measurement workflow that utilizes the contoured mask of TPs-related features \cite{Yang2022}. Banerjee et al. proposed a hybridized, multi-scale feature fusion U-net to extract semantically rich features and fuse multi-scale features \cite{Banerjee2022}. Huang et.al. investigated a joint network for spine segmentation with the interaction of noise-removing work. A selective feature-sharing strategy has been employed to filter out irrelevant features \cite{Huang2022}. However, ultrasound images are significantly prone to noise and speckles, which leads to less-than-ideal segmentation results, reducing the overall precision of UCA measurement. Additionally, manual line drawing on the segmentation region is still required. Due to this, the importance of straightforward landmark identification for line placement is self-evident, regardless of the segmentation performance of vertebral bodies.
	
	Instead of drawing lines within the segmentation region, we approach the measurement of UCA as a pose estimation task, i.e., connecting joint points in a line. Bottom-up approaches are practical for conducting UCA as they first identify individual keypoints on an image before assembling them into objects. OpenPose, a method which utilizes convolutional neural networks (CNNs) and part affinity fields to detect keypoints and estimate human poses, demonstrates excellent performance in multi-person scenarios and can handle occlusion and overlapping body parts \cite{Cao2018}. Another bottom-up approach is the use of associate embeddings \cite{Newell2017}. This method introduces an additional channel in the CNN output, called the “tag map”, which contains an embedding value for each keypoint. The tag map is used to group detected keypoints belonging to the same instance by minimizing the distance between their embedding values. This approach simplifies the process of assembling keypoints into specific objects and has shown promising results in terms of accuracy and efficiency \cite{Li2021, Kreiss2019, Wang2022, Geng2021}. However, parsing the correlation among corresponding landmarks becomes significantly intricate in the context of ultrasound coronal images due to the intrinsic similarity of adjacent vertebral features. We adopt a density-based clustering algorithm to represent both the orientation and positional information of individual vertebral regions. The identified landmarks are easily assembled to form a line using the clustered affinity information.

		\section{Method}
		\begin{algorithm}
			\caption{Clustering Affinity Pipeline}
			\begin{algorithmic}[1]
				\REQUIRE Set of foreground segment points $\mathbf{p}^\mathbf{i}, i=1,2,\ldots,n$ in predicted segmentation map
				\REQUIRE Threshold $\gamma$
				\ENSURE Clustered affinity map $A(p)$
				
				\STATE Calculate neighborhood density for each point:
				\FOR{$i = 1$ to $n$}
				\STATE Define neighborhood of point $\mathbf{p}^\mathbf{i}$ as path $\vec{\mathbf{p}^\mathbf{i}}$ where values are non-zero
				\STATE Estimate density as number of points in neighborhood
				\ENDFOR
				
				\STATE Identify core points:
				\FOR{$i = 1$ to $n$}
				\IF{Number of points in $\vec{\mathbf{p}^\mathbf{i}}$ is at least $\gamma$}
				\STATE Mark $\mathbf{p}^\mathbf{i}$ as core point
				\STATE Mark the points on the path connected to the core point
				\ENDIF
				\ENDFOR
				
				\STATE Assign points to clusters:
				\FOR{$i = 1$ to $n$}
				\IF{$\mathbf{p}^\mathbf{i}$ is core point and the points connected to this point are not in any cluster}
				\STATE Add $\mathbf{p}^\mathbf{i}$ to new cluster
				\ELSE
				\STATE Add $\mathbf{p}^\mathbf{i}$ to same cluster as core point
				\ENDIF
				\ENDFOR
				
				\STATE Sort points in each cluster by x-coordinate:
				\FOR{each cluster}
				\STATE Sort points in cluster by x-coordinate to obtain sequence $S$
				
				\STATE Divide clusters into two subsequences based on leftmost and rightmost points:
				\STATE Calculate centroids $c_l^v$ and $c_r^v$
				\STATE Divide cluster into two subsequences using centroids
				\ENDFOR
				
				\STATE Calculate clustered affinity map:
				\FOR{$i = 1$ to $n$}
				\IF{$\mathbf{p}^\mathbf{i}$ is on vertebra $v$}
				\STATE $A(p) = \frac{c_r^v - c_l^v}{\left \|c_r^v - c_l^v \right \|^2}$
				\ELSE
				\STATE $A(p) = 0$
				\ENDIF
				\ENDFOR
				
				\RETURN Clustered affinity map $A(p)$
			\end{algorithmic}
		\end{algorithm}
	The framework is graphically illustrated in Figure.3. It consists of a feature extraction backbone followed by a landmark detection module and a vertebrae segmentation module. These modules are jointly optimized to regress the orientation and location information of landmarks. The orientation information is represented using the affinity cluster strategy on the predicted segmentation map. The candidate landmarks are grouped to form the lines for UCA measurement based on the clustered affinity map. Specifically, we instantiate the backbone using High-Resolution Network (HRNet) to perform feature extraction \cite{Wang2021}.
	The latent features are then divided into non-overlapping patches, and a positional embedding layer is employed to preserve local continuity and global positional information. A fully connected layer then maps the dimensions of all patches to specific channels for the decoders to compute global dependencies and local context. The landmark decoder comprises a series of transformer blocks that take features as inputs and generate smoothed Gaussian heatmaps as prediction  \cite{Dosovitskiy2021, payer2019integrating}. 
	Let $\mathbf{p^{*}_{l}}, \mathbf{p^{*}_{r}}\in \mathbb{R}^{2}$  be the left and right endpoint of the vertebra respectively. We use an unnormalized Gaussian kernel applied to each point location to produce ground truth heatmaps, which can be denoted as $H^{*}_{(l,r)}(\mathbf{p}) = Gaussian((\mathbf{p^{*}_{l}}, \mathbf{p^{*}_{r}}), \sigma )$, where $H^{*}_{(l,r)}$ is the ground truth Gaussian heatmap and $\sigma$ controls the standard deviation of the Gaussian kernel. We use a mean square error loss to compute the difference between the predicted heatmap and the ground truth heatmap. The heatmap loss function $\ell_{\mathbf{hp}}$ is as follows:
	\begin{equation}
		\centering
		\ell_{\mathbf{hp}} =-\min_{\sigma} \sum_{\mathbf{p}} (\left \|H^{*}_{l, r}(\mathbf{p}, \sigma) - H_{l, r}(\mathbf{p}) \right \|^{2} ) + \left \|\sigma  \right \| _{2}^{2} 
	\end{equation}
	where $H_{(l, r\author{names})} \in \mathbb{R}^{4*H*W}$ is the pixel-wise predicted heatmap representing the thoracic and lumbar landmarks on the left and right sides.
	The regularization term in the L2 norm encourages the values of $\sigma$ to be minimized, while the former objective function favors larger $\sigma$ values. This creates a balance where larger $\sigma$ can result in oversmoothed predictions that may be inaccurate, while smaller sigma values can lead to highly accurate responses but with multiple peaks nearby.
	In addition, we introduce the features in segmentation decoder into landmark decoder. The motivation is to utilize the segmentation information to make point detection more focused on the vertebra region and suppress irrelevant information. Different from the standard multi-head attention, the keys in the region concentration bridge are the fused feature maps from the segmentation branch. Then the query from landmark decoder were computed with all the keys. The attention function can be formulated as follows: 
	\begin{figure}[t]
		\centering
		\includegraphics[height=0.4\textheight, width=0.46\textwidth]{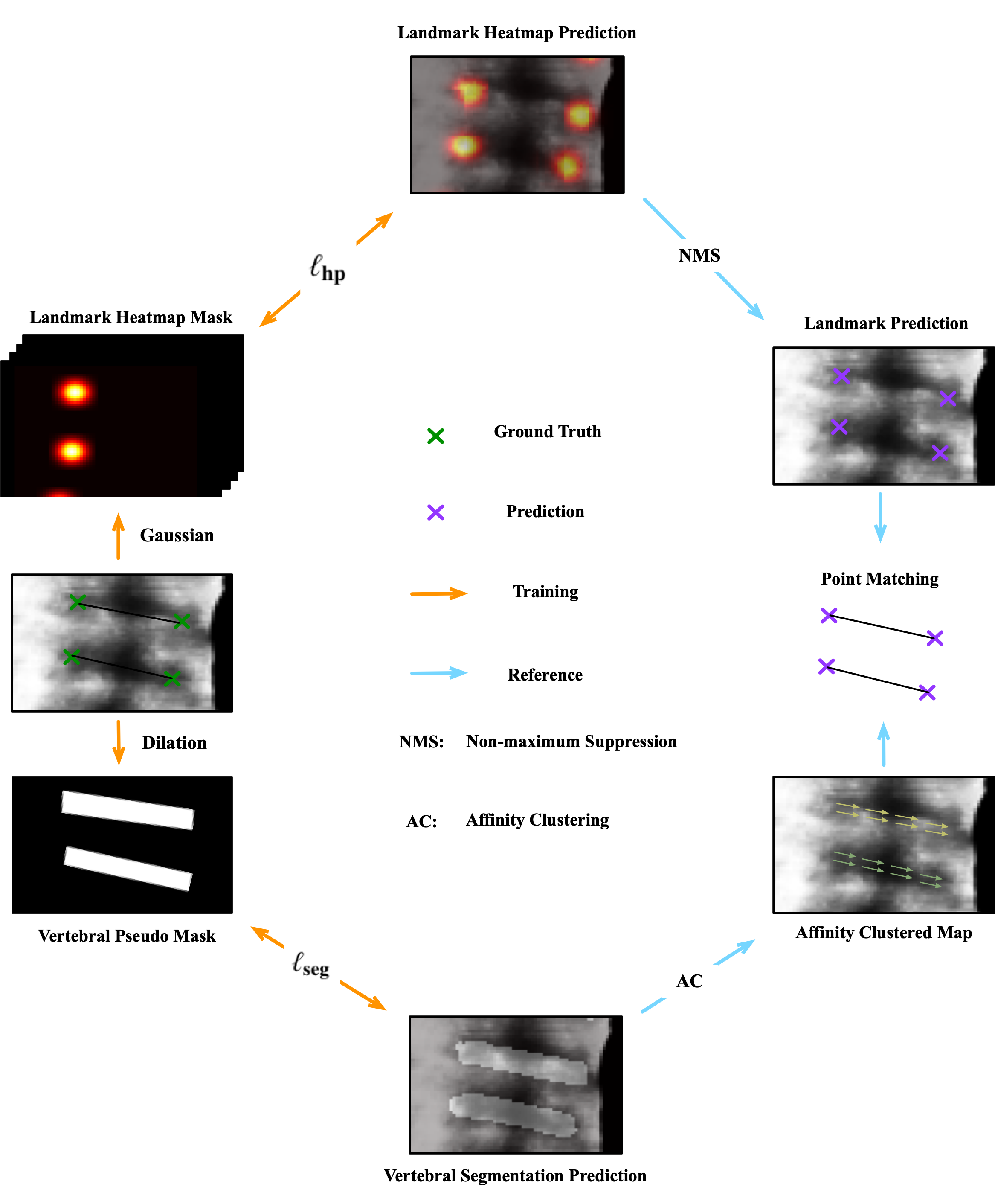}
		\caption{. The pipeline of automatic UCA measurement on stage of training and reference.}
	\end{figure}
	\begin{equation}
		Attention(Q, K_{Seg}, V) = SoftMax(\frac{Q(K_{Seg})^T}{\sqrt{d_{head}}})V
	\end{equation}

	Upon identifying the locations of the landmarks through non-maximum suppression on the heatmap, we perform bipartite matching to associate candidate points, thereby establishing line placement. 
	An affinity map that preserves the association between candidate points is the prerequisite for point matching. We have observed that simultaneously segmenting the vertebra and parsing out the inclination of individual vertebrae yields better performance compared to regressing the affinity field using a neural network.
	This approach enables a subsequent affinity clustering strategy to gather both location and orientation information across the segmented region. 
	We create pseudo-masks $\mathbf{y^*}$ as the ground truth segmentation map, by leveraging ground truth UCA line segments. Specifically, given a set of ground truth line segments $L = {l_1, l_2, ..., l_n}$ representative  of the connections between same vertebral landmarks, we define a fixed-size convolution kernel $K$ and apply a dilation operation to each line segment $l_i \in L$:
	\begin{equation}
		\left(l_{i} \oplus K\right)(x, y)=\bigcup_{(i, j) \in K} l(x-i, y-j)
	\end{equation}
	\begin{equation}
		\mathbf{y^*} = \bigcap\limits_{i=1}^n (l_i \oplus K)
	\end{equation}
	The operation $\oplus$ represents dilation.
	For each pixel location $(x, y)$, the result of $(l_{i} \oplus K)$ is the union of the pixels from $l(x-i, y-j)$ for all offsets $(i, j)$ within the kernel $K$.
	The intersection operation $\bigcap$ is then applied to all these dilated line segments.
	The promising $\mathbf{y^*}$ represents the common areas of dilation across all the ground truth line segments, effectively creating a segmentation map that captures the shared information from these line segments.
	After acquiring the pseudo-masks, the segmentation loss is based on the Dice loss as follows:
	\begin{equation}
		\ell_{\mathbf{seg}\ }=\ 1\ -\ \frac{2\left|\mathbf{y}^\ast\bigcap\mathbf{y}\right|}{\left|\mathbf{y}^\ast\right|\ +\ \left|\mathbf{y}\right|\ }
	\end{equation}
	The training pipeline are illustrated in Figure.4. The total loss $\ell_{total}$ is:
	\begin{equation}
		\ell_{total} = \lambda_{1} \times \ell_{\mathbf{hp}} + \lambda_{2} \times \ell_{\mathbf{seg}}
	\end{equation}
	In our experiment, we empirically set $\lambda_1$ = 1 and $\lambda_2$ = 0.2, respectively. 
	
	\subsection{Vertebral Affinity Clustering}

	The segmentation map not only reveals the position of each vertebra but also illustrates the spine's orientation, setting the stage for subsequent point matching. The clustering affinity pipeline is outlined in Algorithm 1.
	Given the set of foreground segment points $\mathbf{p}^\mathbf{i}, i=1,2,\ldots,n$ in the predicted segmentation map, we calculate the density of its neighborhood for each point. A point's neighborhood is defined as the path $\vec{\mathbf{p}^\mathbf{i}}$ where the values from the neighborhood point to this point are non-zero. Density is estimated by the number of points in the neighborhood. Next, we establish a threshold $\gamma$, where a point is considered a core point if it has at least $\gamma$ other points within its neighborhood. All points on the path are connected to the core point, and then assigned to the same vertebra cluster. If a point is not a core point but is connected to some core points, it is also assigned to the same cluster as that core point. Points that are neither core points nor connected to one are treated as noise points.
	
	Following the clustering procedure, each core point belongs to an individual cluster. For each cluster representative of the potential area of a vertebra $v$, we arrange all points in ascending order based on their x-coordinate to obtain a sequence $S$. Based on the distance of each point in the cluster to the leftmost and rightmost points within the cluster, we divide the cluster into two subsequences and subsequently calculate the centroids. Ultimately, the orientation information of a point $p$ can be represented using a two-dimensional vector map:
				\begin{figure}[t]
			\centering
			\includegraphics[height=0.4\textheight, width=0.35\textwidth]{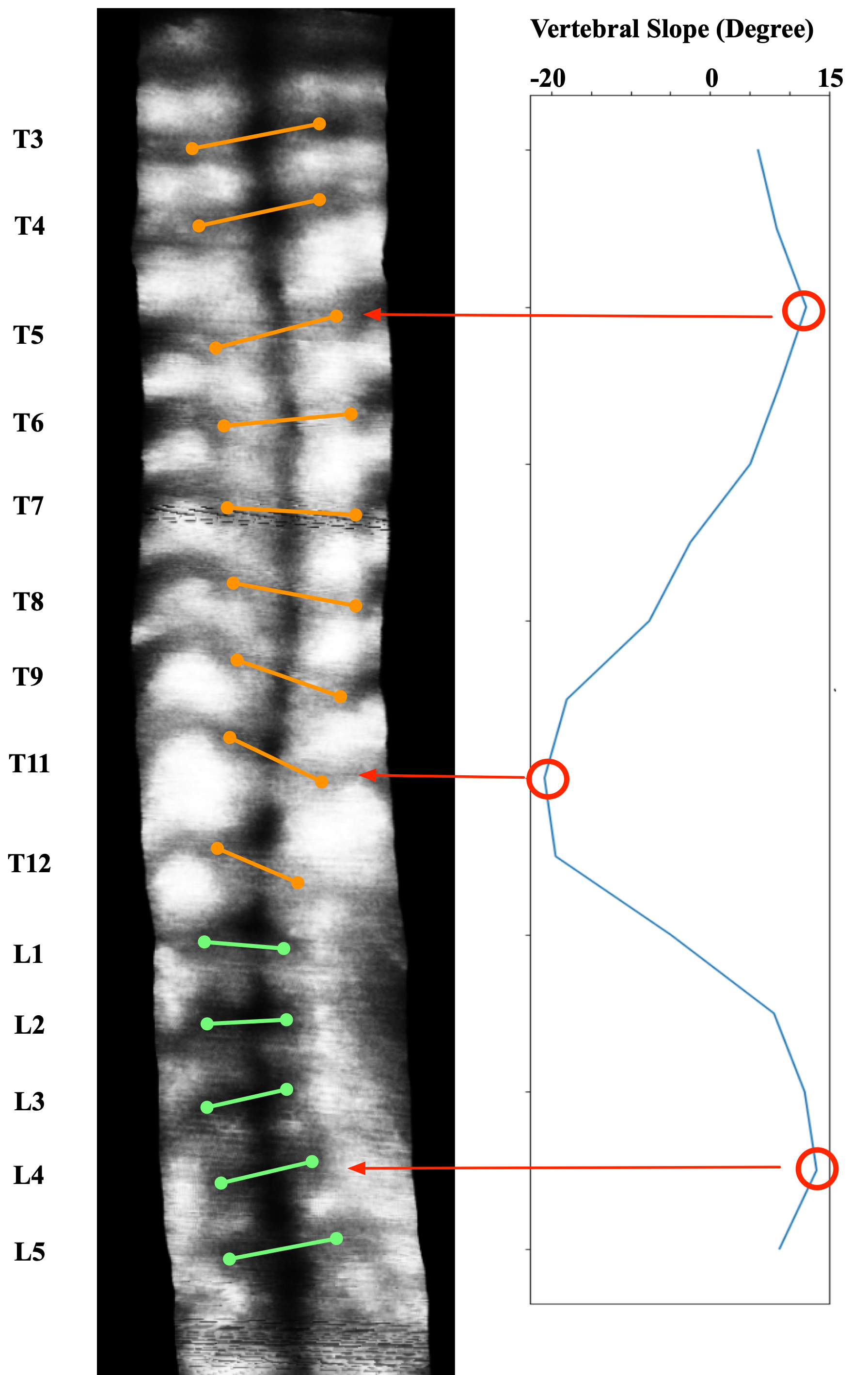}
			\caption{. An example of vertebral-level line detection. The local extreme values are formed the UCA for angle measurement.}
		\end{figure}
\begin{equation}
	A(p) = \begin{cases}
		\frac{c_r^v - c_{l}^{v}}{\left \|c_r^v- c_l^v \right \|^2 }  &  \text{ if  $p$\ on\ vertebrae\ $v$} \\
		0 & \text{ otherwise }
	\end{cases}
\end{equation}
	\begin{table*}
	\begin{center}
		\caption[]{Comparison of different advanced methods on line prediction in the thoracic and lumbar regions.}
		\begin{flushleft}
		\begin{threeparttable}
			\setlength{\tabcolsep}{8mm}{
				\begin{tabular}{c cc cc}
					\toprule[0.2mm]
					\multirow{2}{*}{Method} & \multicolumn{2}{c}{$\mathbf{Thoracic}$}  & \multicolumn{2}{c}{$\mathbf{Lumbar}$}\\
					& AR & AP & AR & AP\\
					\midrule[0.2mm]

					Hourglass$^{*}$ & 0.871 $\pm$ 0.12 & 0.788 $\pm$ 0.16& 0.839 $\pm$ 0.20 &  0.789 $\pm$ 0.22 \\
					HRNet-W32$^{*}$&  0.863 $\pm$ 0.13 &  0.875 $\pm$ 0.13 &  0.842 $\pm$ 0.24  &  0.791 $\pm$ 0.22 \\
					HRNet-W48$^{*}$ &  0.871 $\pm$ 0.12 &  0.875 $\pm$ 0.16&  0.870 $\pm$  0.18&  0.799 $\pm$ 0.21 \\
					HigherHRNet-W32$^{*}$& 0.909  $\pm$ 0.09& 0.941 $\pm$  0.11&  0.874 $\pm$ 0.18  &  0.784 $\pm$ 0.21 \\
					HigherHRNet-W48$^{*}$&0.934 $\pm$ 0.08& 0.928 $\pm$ 0.11& 0.901 $\pm$ 0.14 & 0.858 $\pm$ 0.16 \\
					DEKR& $\mathbf{}$0.945 $\pm$ 0.07 &  0.938 $\pm$ 0.06&  0.902 $\pm$ 0.12 &  0.903 $\pm$ 0.13\\
					CID & 0.915 $\pm$ 0.08&  0.942 $\pm$ 0.06 &  0.843 $\pm$ 0.14 &  0.901 $\pm$ 0.14\\
					Ours &$\mathbf{0.963}$ $\pm$  0.08& $\mathbf{0.950}$ $\pm$  0.08& $\mathbf{0.918}$ $\pm$ 0.10 & $\mathbf{0.907}$ $\pm$ 0.11 \\
					
					\bottomrule[0.2mm]
			\end{tabular}
			\begin{tablenotes}
				\footnotesize
				\item[*] Indicates using associate embedding for point grouping, AP: Average precision, AR: Average recall
			\end{tablenotes}
		}
		\end{threeparttable}
	\end{flushleft}
	\end{center}
\end{table*}
\begin{table}[t]
	\begin{center}
		\caption[]{Ablation study of the size of dilated kernel and segmentation feature transformation.}
		\setlength{\tabcolsep}{0.7mm}{
			\begin{tabular}{c c cc cc}
				\toprule[0.2mm]
				\multirow{2}{*}{FF} & \multirow{2}{*}{Kernel} & \multicolumn{2}{c}{$\mathbf{Thoracic}$}  & \multicolumn{2}{c}{$\mathbf{Lumbar}$}\\
				& & AR & AP & AR & AP\\
				\midrule[0.2mm]
				&1&0.951(0.12)& 0.930(0.10)&0.905(0.09)& 0.891(0.09)\\
				\checkmark&1&0.955(0.10)& 0.940(0.10)&0.910(0.10)& 0.890(0.09)\\
				&3&0.956(0.10)& 0.941(0.10)& 0.911(0.10) & 0.906(0.11)\\
				\checkmark&3&$\mathbf{0.963}$(0.08)& $\mathbf{0.950}$(0.08)& $\mathbf{0.918}$(0.10) & 0.907(0.11)\\
				&5&0.958(0.10)& 0.940(0.09)&0.917(0.10)& $\mathbf{0.910}$(0.09)\\
				\checkmark&5&0.961(0.09)& 0.947(0.08)&0.915(0.11)& 0.906(0.08)\\

				
				\bottomrule[0.2mm]
		\end{tabular}}
		\label{Table2}
	\end{center}
	
\end{table}

	Here, $A(p)$ represents the clustered affinity map. If the point is located on vertebra $v$, the value is a unit vector from the left centroid $\mathbf{c}_\mathbf{l}^\mathbf{v}$ to the right centroid $\mathbf{c}_\mathbf{r}^\mathbf{v}$, indicating the direction of vertebra $v$. For points outside the vertebra cluster, the value is zero.
	
	\subsection{Vertebrae Parsing}
	For both thoracic and lumbar regions, we have obtained several candidate points after using non-maximum suppression operation on the landmark heatmaps. These candidates deﬁne a large set of potential lines representative of the vertebrae.
	Specifically, we have a set of points distributed on the left and right sides of the spinous process profile, where $S^l = {d_i,\ i\in{1\ ...\ I}}$ and $S^r = {d_j,\ j\in{1\ ...\ J}}$, where $I$ and $J$ are the number of detected points on the left and right side, respectively. 
	We then perform the line integral computation along the path of two candidates $\vec{A}(\ast)$ on the clustered affinity map.
	\begin{equation}
		c_{ij} =\int_{u=0}^{u=1}{\vec{A}(\left(1-u\right)d_i + ud_j )du}
	\end{equation}
	The $c_{ij}$  indicates the confidence of whether the $d_i$ and $d_j$  are connected to form the line. The optimal matching of candidates is achieved with the Hungarian algorithm to acquire all the vertebrae connections \cite{Kuhn1955}. After obtaining the optimal matching, we can filter out low-confidence matches based on  the criteria that confidence significantly lower than the average of finished matches are ignored, as they might intersect with other line segments or be less reliable. 
	The line of interest for UCA measurements is based on the horizontal slope between each pair of line segments and their adjacent counterparts to identify local extrema (peaks and valleys) (Figure 5). Additionally, if the absolute value of the angle formed by the detected uppermost or lowermost vertebral bodies in the global space and the most inclined vertebral body obtained in the local space exceeds 10 degrees, the UCA is calculated. Consequently, all detected vertebral bodies are considered for UCA computations.

	\section{Experiments}
	\subsection{Materials}
	All participants were recruited from the Department of Orthopedics and Traumatology of The Chinese University of Hong Kong. Informed consents are obtained before the scanning session. Patients with a Cobb angle greater than 60$^\circ$ and BMI indices above 25.0 kg/m² were excluded. 
	VPI images were acquired by two 3D ultrasound imaging systems: Scolioscan 801 and Scolioscan Air  \cite{Lai2021}. 
	For model development, three experts with more than 5 years of ultrasound experiments manually annotated the line between the spinal feature points on both sides of the spinous process profile as ground truth. A total of 1212 cases were included, with 970 cases used to train the model and 242 used for the in-house validation dataset to evaluate the performance. 386 prospective cases with biplanar radiographs were used to test the performance of model after the model was developed. The Cobb angles were measured by two radiograph experts.
	
	The model was implemented based on PyTorch and trained on a 48GB NVIDIA RTX A6000 GPU. The data augmentation included random horizontal flipping, rotation ranging from -30 to 30 degrees, brightness, and contrast transformation. We resized the input images into 256 $\times$ 512, keeping the aspect ratio. We empirically set the $\gamma$ = 10 to filter the noise segment region. The initial learning rate was $1e^{-5}$ and an Adam optimizer with a momentum of 0.9 was employed for model development. 
					\begin{table}[t]
		\begin{center}
			\caption[]{Ablation study of different point grouping strategies.}
			\setlength{\tabcolsep}{0.3mm}{
				\begin{tabular}{c cc cc}
					\toprule[0.2mm]
					\multirow{2}{*}{Grouping} &\multicolumn{2}{c}{$\mathbf{Thoracic}$}  & \multicolumn{2}{c}{$\mathbf{Lumbar}$}\\
					&AR & AP & AR & AP\\
					\midrule[0.2mm]
					AE&0.922 (0.11)& 0.910 (0.09)&0.875 (0.11)& 0.879 (0.11)\\
					PAF&0.931 (0.10)& 0.904 (0.09)&0.885 (0.11)& 0.890 (0.12)\\
					AC&$\mathbf{0.963}$ (0.08)& $\mathbf{0.950}$ (0.08)& $\mathbf{0.918}$ (0.10) & $\mathbf{0.907}$ (0.11)\\
					\bottomrule[0.2mm]
			\end{tabular}}
			\label{Table3}
		\end{center}
		
	\end{table}
		\begin{figure}[t]
	\centering
	\includegraphics[height=0.33\textheight, width=0.22\textwidth]{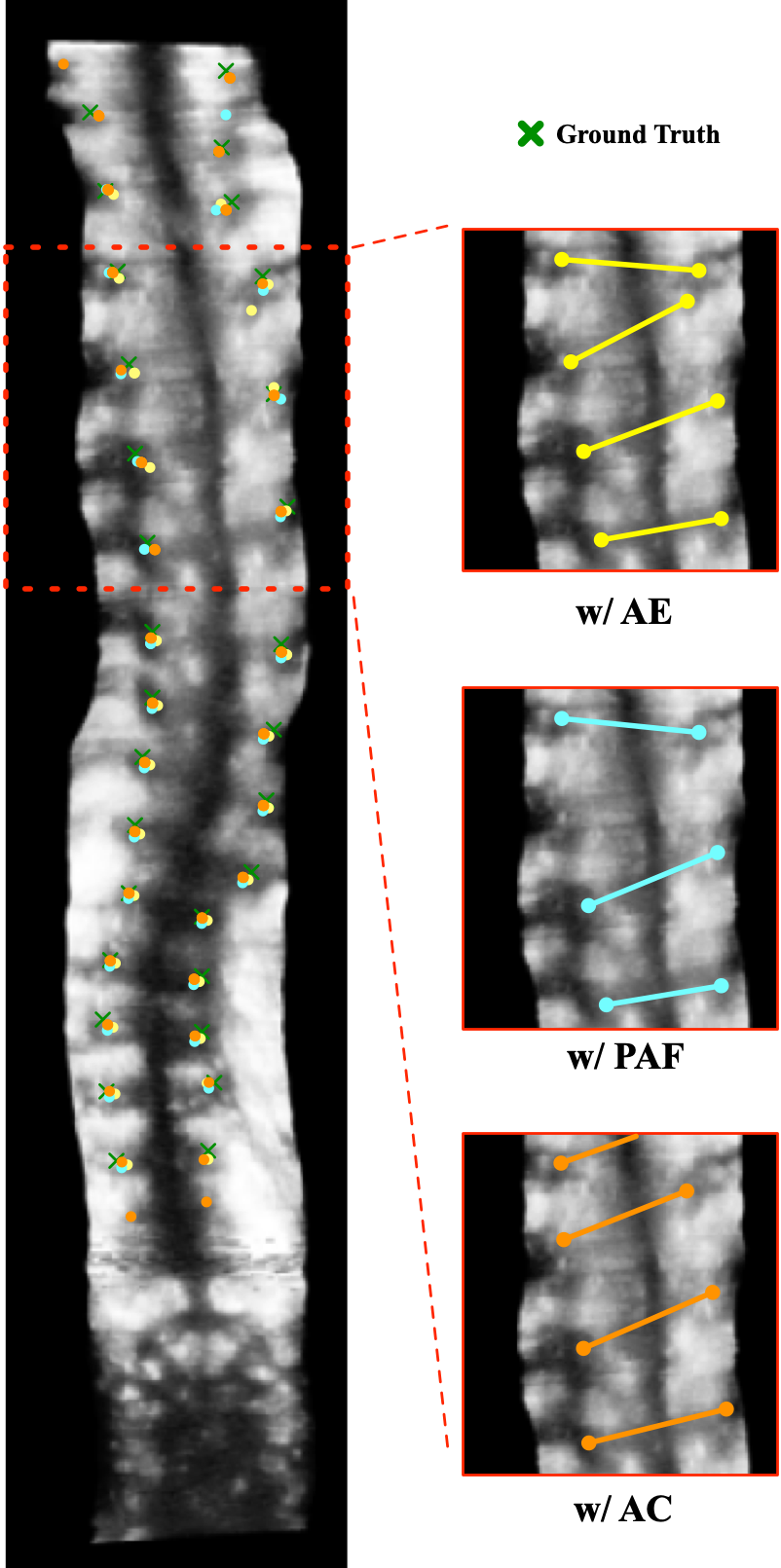}
	\caption{. Sample image in ablation study of using different point grouping strategy. The similarity in the features of vertebrae could result in incorrect line connections. }
\end{figure}
		\begin{figure}[t]
	\centering
	\includegraphics[height=0.4\textheight, width=0.4\textwidth]{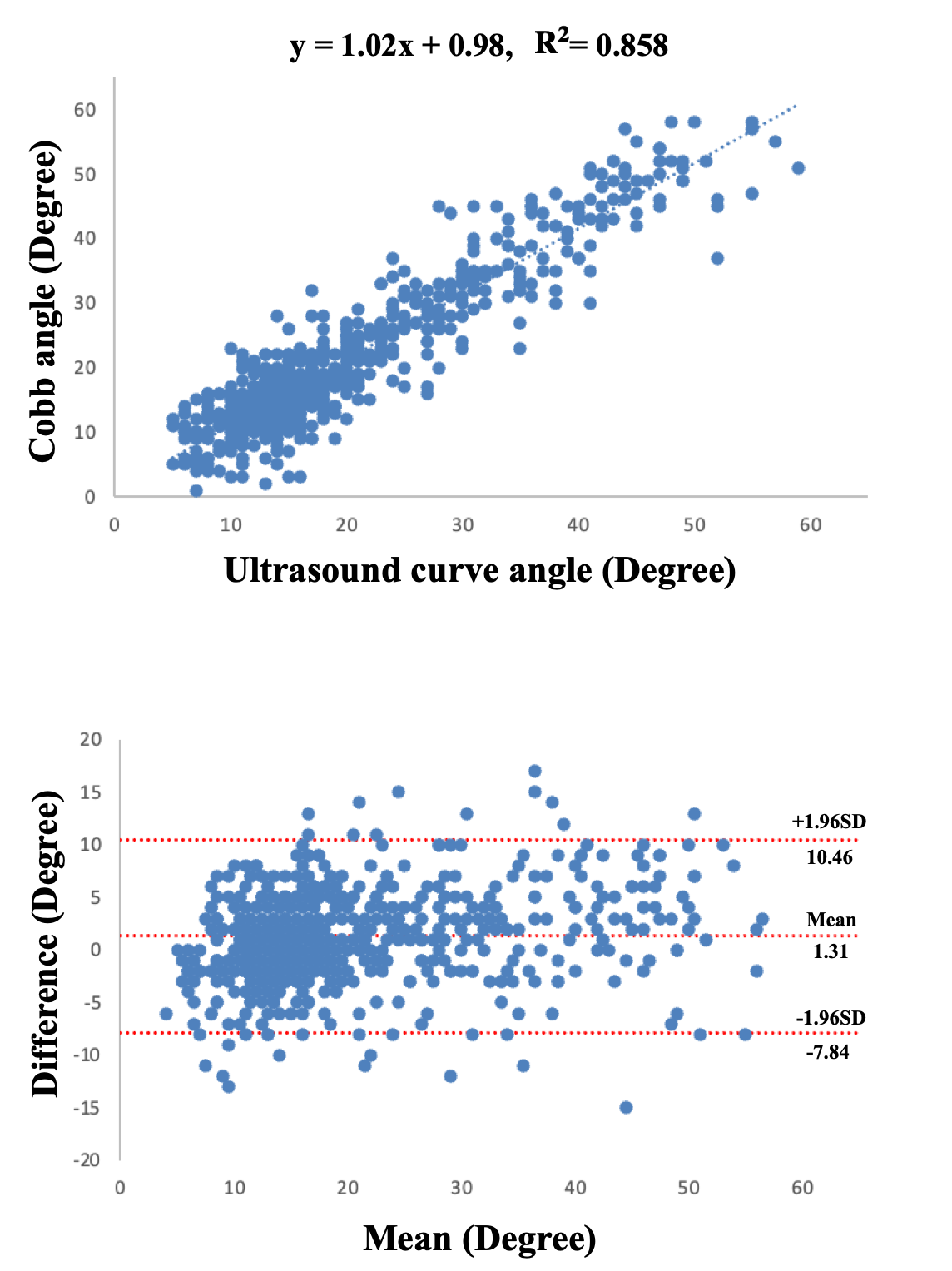}
	\caption{. Linear regression analysis and Bland-Altman plots of predicted UCA on the test data.}
	\label{linear_altman}
\end{figure}
	\subsection{Evaluation Metrics}
	The performance evaluation of the model was conducted using vertebral-level line prediction and UCA measurement. Traditional metrics such as Mean Euclidean Distance (MED) and Mean Manhattan Distance (MMD) are often used to assess the disparity between predicted points and ground truth \cite{Meng2022, Meng2023}. However, these metrics may not fully account for the possibility of redundant or missing predicted lines.
	To provide a more comprehensive evaluation of the line prediction performance, we integrate the point-based metrics with the concept of Endpoint Distance Error (EDE) as follows: 
	\begin{equation}
		{EDE}=\frac{\sum_{l,r} e x p\left(-d_{l,r}/s\right)}{2}.
	\end{equation}
	Here $d_{l,r}$ stands for the Euclidean Distance of the left and right endpoint of the predicted line with its corresponding ground truth. We set $s$ equal to 100, which is a scaling factor. The EDE measures the localization accuracy of a single line, whereases correct prediction is defined as the distance between the predicted endpoint and the ground truth less than 3.5mm (EDE $>$ 0.5) according to the posterior vertebral body heights obtained from a study using the human cadaver  \cite{kunkel2011}. The perfect prediction would yield an EDE of 1. We define the average precision (AP) and average recall (AR) scores as the ratio of corrected lines to the total ground truth and the ratio of corrected lines to the total of predicted lines in the scan respectively. These metrics consider both the redundancy and absence of predicted lines. To further verify the validity of the UCA measurement, we used linear regression and Bland–Altman analysis to investigate the agreement between predicted UCA and Cobb angle.

	\subsection{Comparison with Advanced Networks}

The performance of line prediction is estimated by comparing it with other advanced methods, including the Hourglass \cite{Newell2016}, HRNet \cite{Wang2021}, HigherHRNe \cite{Cheng2020}, CID \cite{Wang2022}, and DEKR \cite{Geng2021}. We reimplement the methods according to the mmPose\footnote{https://github.com/open-mmlab/mmpose}. For baselines \cite{Newell2016, Wang2021, Cheng2020}, we assemble detected keypoints whose tags with small $L_2$ distance into line by using associative embedding. Table.I summarizes the comparison results. Based on the results, it is evident that our proposed method achieved superior performance in terms of precision and recall. This implies that the predicted line segments are accurate within the specified range, and the occurrence of redundant line segments is minimal. We attribute this notable performance difference to the limitations of other methods, particularly in their regression of association between keypoints. These methods fail to effectively address the challenge of erroneous landmark connections that occur due to the structural similarity between the current vertebral body and its neighboring vertebrae. Consequently, the keypoint parsing process generates numerous intersecting line segments, leading to a significant reduction in accuracy.
	Furthermore, relying exclusively on the Gaussian heatmap generated from the labels leads to inadequate performance, particularly in accurately detecting all landmarks in the presence of image blurriness on the vertebral region.
	In contrast, our proposed method benefits from segmentation supervision, enabling precise landmark prediction on both sides of the spine. To address the ambiguity in certain regions predicted by the model, we employ a clustering strategy. This strategy effectively reduces the generation of redundant line segments by classifying these ambiguous areas as noise regions. This approach prevents their impact on the identification of vertebral regions.

	\renewcommand{\dblfloatpagefraction}{.8}
	\begin{figure*}
		\centering
		\includegraphics[height=0.4\textheight, width=0.8\textwidth]{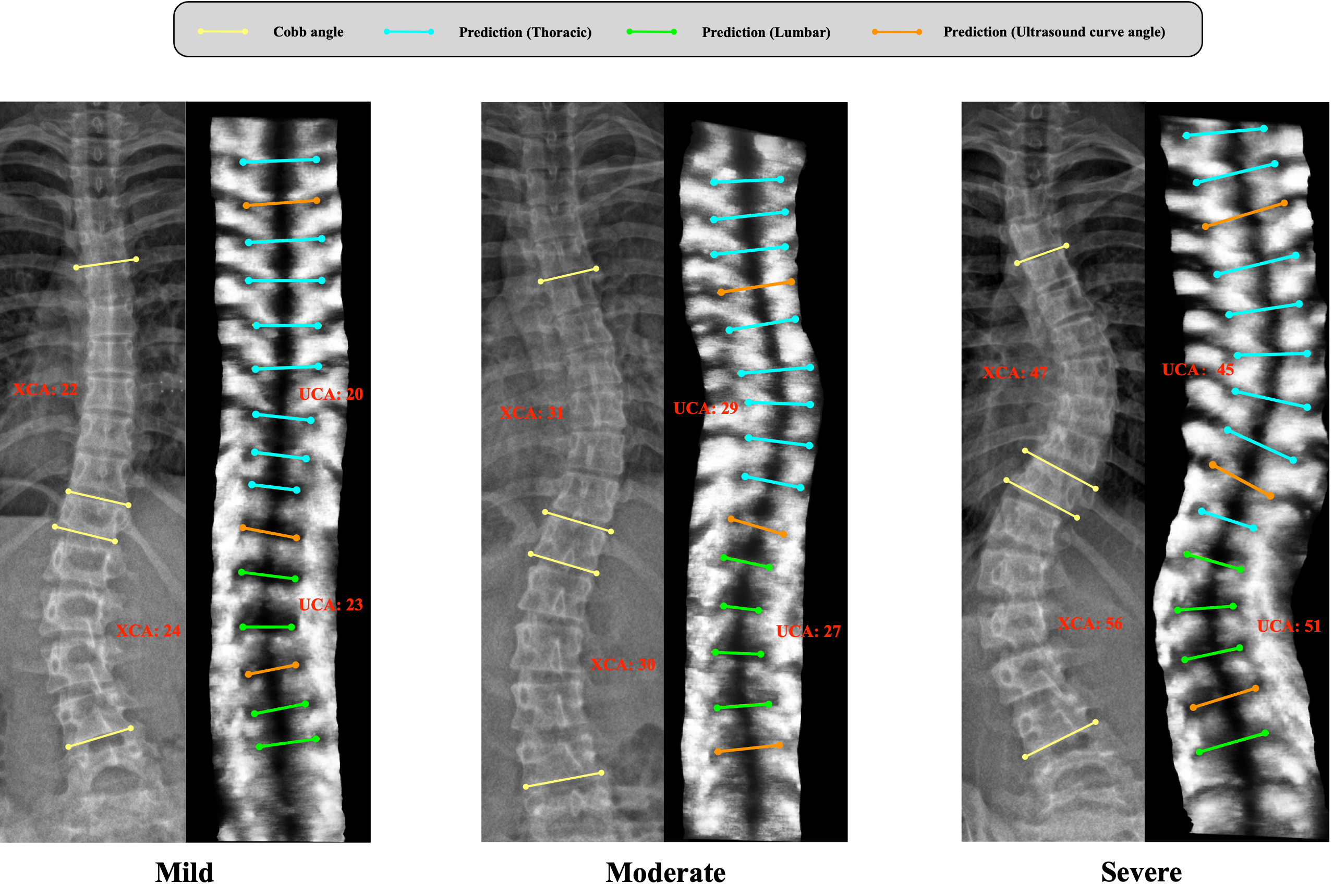}
		\caption{. Visual comparison between Ultrasound Curve Angle (UCA) and X-ray Cobb angle (XCA) in different degrees of scoliosis. }
		\label{visualized}
	\end{figure*}

	\subsection{Ablation Studies}
	This section aims to verify the effectiveness of the proposed components, including the affinity clustering (AC) strategy for point grouping and different size of dilated kernel for vertebral region discrimination. 
	To assess the contribution of our proposed components for landmark detection, we chose different size of dilated kernel to create the mask of vertebral region. We also investigate the effectiveness of feature transformation from segmentation to the landmark detection encoder. The comparison results are shown in Table.II. We observe that the AR and AP are the most optimal when the kernel size is set to 3. 
	A too large, dilated kernel can lead to overlapping spatial position predictions of different vertebrae, resulting in a biased affinity representation towards the direction of the upper and lower vertebrae. Conversely, a too small, dilated kernel makes model convergence more challenging, with an increased likelihood of predicting regions outside the vertebrae. On the other hand, experimental results suggest that introducing representations of vertebral regions can aid in landmark localization, especially when one side's shadow of spinous process are unclear or missing due to the discontinuity in the scanning process.

We investigate the performance of two bottom-up strategies for keypoint grouping, which are associate embedding (AE) and part affinity field (PAF). We modify the output of the head of segmentation to predict the PAF and the embedding heatmap for each candidate, respectively. The results are shown in Table.III. We visualize the predicted landmarks from different grouping strategies with its corresponding line connection in Figure.6. The model performs point grouping based on the part affinity field, but it is more powerful in ultrasound coronal images. Different from directly regressing the affinity field of keypoints, clustered affinity information from segmented spines performed more efficiently for the following reasons. Firstly, spinal structures do not exhibit feature overlap, simplifying the prediction of location information across the region of a vertebra.  Secondly, we notice that the prediction of affinity or embedding heatmap from the neural network is affected by the similar profile of adjacent vertebrae. This results in detected keypoints that tend to be connected to adjacent vertebrae rather than the exact vertebra where the landmark is located.

	\subsection{Comparison with Cobb Angle}
	In this section, we evaluate the correlation between automatic ultrasound curve angle and X-ray Cobb angle. The linear regression analysis and Bland-Altman plots are shown in Figure.7.  Figure.8 visualizes the result between predicted UCAs and Cobb angles for the same patients. The results reveal a strong correlation between the automated UCA and the Cobb angle, evidenced by an R$^2$ value of 0.858. The study result closely aligns with the previously reported results of the comparison between manual UCA and Cobb angle (R$^2$=0.888) \cite{Lee2021}. The Bland-Altman plots indicate an overall mean difference of 1.31 degrees, exhibiting good agreement between the predicted UCA and the Cobb angle. Seventy-six percent of UCA (441 out of 580 curves) exhibit a difference within 5° when compared to the Cobb angle. The scaling factor, derived from the linear equation, is determined to be 1.02, indicating a great agreement between automated UCA and the Cobb angle. We observed that the vertebral bodies contributing to the angle measurement may not align consistently between ultrasound and X-ray imaging. This discrepancy arises from the different approaches used in the two modalities. In X-ray, the apex position is first determined, and then line segments are selected on either side of the apex. In contrast, we directly calculate the inclination of each vertebral body in ultrasound, thereby avoiding the potential occurrence of adjacent vertebral bodies being more inclined than the measured vertebral body. Our approach achieves a more accurate and reliable assessment of the vertebral inclination.

	\section{Discussion}
Previous methods have relied on features of the spinous process for assessing spinal curvature \cite{Huang2023, ungi2020automatic, brignol2020automatic, chen2024development}. However, this method has limitation. Axial vertebral rotation, which is common in mild and moderate scoliosis, can lead to an underestimation of spinal curvature when using spinous process. Therefore, transverse process for spinal curvature measurement is closer to the actual condition of the patient. To this end, we achieve the first fully automatic measurement of scoliosis in ultrasound, based on transverse process. Many existing methods for automatic detection of scoliosis using ultrasound rely heavily on accurate vertebra segmentation \cite{Huang2022, wong2022convolutional, Banerjee2022}. However, these methods can be affected by artifacts in ultrasound images, which may arise due to the discontinuity in probe movement during scanning. In addition, due to the similarity of vertebrae features in ultrasound images, previous landmark-based algorithms struggle to decouple the correlation of each detected key point. As a result, vertebrae curvature still requires manual connection of the most tilted key points detected, making fully automatic end-to-end scoliosis analysis unachievable.  

In contrast, our method employs a bottom-up strategy to perform affinity clustering on the segmentation masks corresponding to each detected key point of the vertebrae. This allows automatic pairing of key points for each vertebra, thereby calculating the slope of each vertebra. Our method was compared with other state-of-the-art detection networks and achieved a correlation of R$^2$=0.835 between automatic UCA and XCA. Previous methods for line detection faced challenges in distinguishing the spatial correlation between adjacent vertebrae due to their similarity in ultrasound images, often leading to connections between different vertebrae landmarks. By leveraging prior knowledge of vertebrae segmentation, our method effectively captures the direction of each vertebra. We also demonstrated a strong correlation with the gold standard of Cobb angle measurement. Notably, our method does not rely on identifying the apex due to the invisibility of the upper and lower edges of vertebrae in ultrasound images; instead, it directly compares the slope of each vertebra to obtain UCA. As an automatic measurement, our method eliminates interobserver variability. It is worth noting that our approach, based on keypoint matching, can be easily extended to the automatic measurement of Cobb angles. Additionally, our method also supports vertebral-level analysis, making it potentially useful for surgical navigation and treatment monitoring \cite{zhang2021flexible, gueziri2020state}.

However, notable angle deviations are observed in cases with poor image quality, primarily due to uncertainties in landmark detection. This suboptimal image quality could potentially be attributed to insufficient contact between the probe and the skin during ultrasound scanning. In addition, the VPI images are generated based on the average skin-to-laminae distances, which can vary among individuals. Consequently, the generated VPI images may not optimally visualize all vertebral features necessary for UCA measurement. This issue also impedes our method's ability to accurately distinguish between the thoracic and lumbar regions in the ultrasound VPI images, a crucial step in the automatic UCA process.
This is because thoracic and lumbar UCAs are computed using different anatomical features derived from the VPI images. Our current approach involves identifying the last pair of ribs on the 12th vertebra to separate the thoracic and lumbar regions, thereby enabling the application of different strategies for assigning UCA lines. However, VPI images generated based on average skin-to-laminae distances may not visualize the 12th ribs. In future studies, we plan to use a gel pad to minimize the likelihood of insufficient contact between the skin and the probe. Additionally, we aim to incorporate information about the ribs into the model to improve the accuracy of distinguishing between the thoracic and lumbar regions. It is well noted that the primary targets of our ultrasound protocol are those preoperative cases, therefore we did not recruit subjects with large Cobb angles. In addition, the 7.5MHz probe used in this study may not be optimal for acquiring good-quality US images, so high BMI subjects are not included. In future studies, ultrasound probes with lower frequencies will be adopted to investigate the feasibility of using these probes on high BMI subjects.

\section{Conclusion}
	We have proposed a framework for automatic ultrasound curve angle measurement, which identifies the potential landmarks and performs line delineation. Our method addresses the challenge of localizing points to perform line delineation automatically. This is meaningful in the clinical setting because the manual process of drawing lines is both time-consuming and operator-dependent. Different from previous segmentation-based networks on volume data or 2D B-mode images, our approach does not rely on the accurate annotation of vertebral structures. In addition, transferring segmentation features into landmark detection allows the model to focus more on specific target areas, making the prediction of the precise location of potential landmarks accurate. Beyond angle measurement, our approach supports vertebral-level analysis, providing a comprehensive understanding of spinal morphology. The superior performance of our method compared to other advanced methods indicates the effectiveness of the proposed network. Furthermore, experiments on a VPI image dataset with radiographs demonstrated the reliability of the proposed network. With minimal operator interaction and skills required, clinicians can efficiently acquire the angle from ultrasound coronal images, eliminating intra-rater and inter-rater operator variation. This holds great potential for replacing manual UCA measurements.
	
\section*{Acknowledgments}
	This study was partially support by The Research Grant Council of Hong Kong (R5017-18).
	
	\bibliographystyle{elsarticle-num}
	\bibliography{Automatic_Ultraosound_Curve_Angle.bib}

\begin{thebibliography}{10}
\expandafter\ifx\csname url\endcsname\relax
  \def\url#1{\texttt{#1}}\fi
\expandafter\ifx\csname urlprefix\endcsname\relax\def\urlprefix{URL }\fi
\expandafter\ifx\csname href\endcsname\relax
  \def\href#1#2{#2} \def\path#1{#1}\fi

\bibitem{Konieczny2013}
M.~R. Konieczny, H.~Senyurt, R.~Krauspe, Epidemiology of adolescent idiopathic scoliosis (2013).
\newblock \href {https://doi.org/10.1007/s11832-012-0457-4} {\path{doi:10.1007/s11832-012-0457-4}}.

\bibitem{Simony2016}
A.~Simony, E.~J. Hansen, S.~B. Christensen, L.~Y. Carreon, M.~O. Andersen, Incidence of cancer in adolescent idiopathic scoliosis patients treated 25 years previously, European Spine Journal 25 (2016).
\newblock \href {https://doi.org/10.1007/s00586-016-4747-2} {\path{doi:10.1007/s00586-016-4747-2}}.

\bibitem{Himmetoglu2015}
S.~Himmetoglu, M.~F. Guven, N.~Bilsel, Y.~Dincer, Dna damage in children with scoliosis following x-ray exposure, Minerva pediatrica 67 (2015).

\bibitem{McArthur2015}
N.~McArthur, D.~P. Conlan, J.~R. Crawford, Radiation exposure during scoliosis surgery: A prospective study, Spine Journal 15 (2015).
\newblock \href {https://doi.org/10.1016/j.spinee.2014.12.149} {\path{doi:10.1016/j.spinee.2014.12.149}}.

\bibitem{jeon2018combination}
C.-H. Jeon, K.-S. Kwack, S.~Park, H.-D. Lee, N.-S. Chung, Combination of whole-spine lateral radiograph and lateral scanogram in the assessment of global sagittal balance, The Spine Journal 18~(2) (2018) 255--260.

\bibitem{Cheung2015_VPI}
C.~W.~J. Cheung, G.~Q. Zhou, S.~Y. Law, T.~M. Mak, K.~L. Lai, Y.~P. Zheng, Ultrasound volume projection imaging for assessment of scoliosis, IEEE Transactions on Medical Imaging 34 (2015) 1760--1768.
\newblock \href {https://doi.org/10.1109/TMI.2015.2390233} {\path{doi:10.1109/TMI.2015.2390233}}.

\bibitem{Lee2021}
T.~T.~Y. Lee, K.~K.~L. Lai, J.~C.~Y. Cheng, R.~M. Castelein, T.~P. Lam, Y.~P. Zheng, 3d ultrasound imaging provides reliable angle measurement with validity comparable to x-ray in patients with adolescent idiopathic scoliosis, Journal of Orthopaedic Translation 29 (2021) 51--59.
\newblock \href {https://doi.org/10.1016/j.jot.2021.04.007} {\path{doi:10.1016/j.jot.2021.04.007}}.

\bibitem{Cheung2015}
C.~W.~J. Cheung, G.~Q. Zhou, S.~Y. Law, K.~L. Lai, W.~W. Jiang, Y.~P. Zheng, Freehand three-dimensional ultrasound system for assessment of scoliosis, Journal of Orthopaedic Translation 3 (2015).
\newblock \href {https://doi.org/10.1016/j.jot.2015.06.001} {\path{doi:10.1016/j.jot.2015.06.001}}.

\bibitem{Chen2013}
W.~Chen, E.~H. Lou, P.~Q. Zhang, L.~H. Le, D.~Hill, Reliability of assessing the coronal curvature of children with scoliosis by using ultrasound images, Journal of Children's Orthopaedics 7 (2013).
\newblock \href {https://doi.org/10.1007/s11832-013-0539-y} {\path{doi:10.1007/s11832-013-0539-y}}.

\bibitem{Zhou2020}
G.~Q. Zhou, D.~S. Li, P.~Zhou, W.~W. Jiang, Y.~P. Zheng, Automating spine curvature measurement in volumetric ultrasound via adaptive phase features, Ultrasound in Medicine and Biology 46 (2020).
\newblock \href {https://doi.org/10.1016/j.ultrasmedbio.2019.11.012} {\path{doi:10.1016/j.ultrasmedbio.2019.11.012}}.

\bibitem{Dang2019}
Q.~Dang, J.~Yin, B.~Wang, W.~Zheng, Deep learning based 2d human pose estimation: A survey, Tsinghua Science and Technology 24 (2019).
\newblock \href {https://doi.org/10.26599/TST.2018.9010100} {\path{doi:10.26599/TST.2018.9010100}}.

\bibitem{Wang2021_survey}
C.~Wang, F.~Zhang, S.~S. Ge, A comprehensive survey on 2d multi-person pose estimation methods, Engineering Applications of Artificial Intelligence 102 (2021).
\newblock \href {https://doi.org/10.1016/j.engappai.2021.104260} {\path{doi:10.1016/j.engappai.2021.104260}}.

\bibitem{Zheng2016}
Y.~P. Zheng, T.~T.~Y. Lee, K.~K.~L. Lai, B.~H.~K. Yip, G.~Q. Zhou, W.~W. Jiang, J.~C.~W. Cheung, M.~S. Wong, B.~K.~W. Ng, J.~C.~Y. Cheng, T.~P. Lam, A reliability and validity study for scolioscan: A radiation-free scoliosis assessment system using 3d ultrasound imaging, Scoliosis and Spinal Disorders 11 (2016).
\newblock \href {https://doi.org/10.1186/s13013-016-0074-y} {\path{doi:10.1186/s13013-016-0074-y}}.

\bibitem{Wong2019}
Y.~shun Wong, K.~K. lee Lai, Y.~ping Zheng, L.~L. ning Wong, B.~K. wah Ng, A.~L. hang Hung, B.~H. kei Yip, W.~C. wing Chu, A.~W. hung Ng, Y.~Qiu, J.~C. yiu Cheng, T.~ping Lam, Is radiation-free ultrasound accurate for quantitative assessment of spinal deformity in idiopathic scoliosis (is): A detailed analysis with eos radiography on 952 patients, Ultrasound in Medicine and Biology 45 (2019).
\newblock \href {https://doi.org/10.1016/j.ultrasmedbio.2019.07.006} {\path{doi:10.1016/j.ultrasmedbio.2019.07.006}}.

\bibitem{Huang2023}
Y.~Huang, J.~Jiao, J.~Yu, Y.~Zheng, Y.~Wang, Si-mspdnet: A multiscale siamese network with parallel partial decoders for the 3-d measurement of spines in 3d ultrasonic images, Computerized Medical Imaging and Graphics 108 (2023).
\newblock \href {https://doi.org/10.1016/j.compmedimag.2023.102262} {\path{doi:10.1016/j.compmedimag.2023.102262}}.

\bibitem{Yang2022}
D.~Yang, T.~T.~Y. Lee, K.~K.~L. Lai, T.~P. Lam, W.~C.~W. Chu, R.~M. Castelein, J.~C.~Y. Cheng, Y.~P. Zheng, Semi-automatic ultrasound curve angle measurement for adolescent idiopathic scoliosis, Spine Deformity 10 (2022) 351--359.
\newblock \href {https://doi.org/10.1007/s43390-021-00421-4} {\path{doi:10.1007/s43390-021-00421-4}}.

\bibitem{Banerjee2022}
S.~Banerjee, J.~Lyu, Z.~Huang, F.~H. Leung, T.~Lee, D.~Yang, S.~Su, Y.~Zheng, S.~H. Ling, Ultrasound spine image segmentation using multi-scale feature fusion skip-inception u-net (siu-net), Biocybernetics and Biomedical Engineering 42 (2022) 341--361.
\newblock \href {https://doi.org/10.1016/j.bbe.2022.02.011} {\path{doi:10.1016/j.bbe.2022.02.011}}.

\bibitem{Huang2022}
Z.~Huang, R.~Zhao, F.~H. Leung, S.~Banerjee, T.~T.~Y. Lee, D.~Yang, D.~P. Lun, K.~M. Lam, Y.~P. Zheng, S.~H. Ling, Joint spine segmentation and noise removal from ultrasound volume projection images with selective feature sharing, IEEE Transactions on Medical Imaging 41 (2022) 1610--1624.
\newblock \href {https://doi.org/10.1109/TMI.2022.3143953} {\path{doi:10.1109/TMI.2022.3143953}}.

\bibitem{Cao2018}
Z.~Cao, G.~Hidalgo, T.~Simon, S.-E. Wei, Y.~Sheikh, \href{http://arxiv.org/abs/1812.08008}{Openpose: Realtime multi-person 2d pose estimation using part affinity fields} (12 2018).
\newline\urlprefix\url{http://arxiv.org/abs/1812.08008}

\bibitem{Newell2017}
A.~Newell, Z.~Huang, J.~Deng, Associative embedding: End-to-end learning for joint detection and grouping, Vol. 2017-December, 2017.

\bibitem{Li2021}
Y.~Li, S.~Zhang, Z.~Wang, S.~Yang, W.~Yang, S.~T. Xia, E.~Zhou, Tokenpose: Learning keypoint tokens for human pose estimation, 2021.
\newblock \href {https://doi.org/10.1109/ICCV48922.2021.01112} {\path{doi:10.1109/ICCV48922.2021.01112}}.

\bibitem{Kreiss2019}
S.~Kreiss, L.~Bertoni, A.~Alahi, Pifpaf: Composite fields for human pose estimation, Vol. 2019-June, 2019.
\newblock \href {https://doi.org/10.1109/CVPR.2019.01225} {\path{doi:10.1109/CVPR.2019.01225}}.

\bibitem{Wang2022}
D.~Wang, S.~Zhang, Contextual instance decoupling for robust multi-person pose estimation, Vol. 2022-June, 2022.
\newblock \href {https://doi.org/10.1109/CVPR52688.2022.01078} {\path{doi:10.1109/CVPR52688.2022.01078}}.

\bibitem{Geng2021}
Z.~Geng, K.~Sun, B.~Xiao, Z.~Zhang, J.~Wang, Bottom-up human pose estimation via disentangled keypoint regression, 2021.
\newblock \href {https://doi.org/10.1109/CVPR46437.2021.01444} {\path{doi:10.1109/CVPR46437.2021.01444}}.

\bibitem{Wang2021}
J.~Wang, K.~Sun, T.~Cheng, B.~Jiang, C.~Deng, Y.~Zhao, D.~Liu, Y.~Mu, M.~Tan, X.~Wang, W.~Liu, B.~Xiao, Deep high-resolution representation learning for visual recognition, IEEE Transactions on Pattern Analysis and Machine Intelligence 43 (2021).
\newblock \href {https://doi.org/10.1109/TPAMI.2020.2983686} {\path{doi:10.1109/TPAMI.2020.2983686}}.

\bibitem{Dosovitskiy2021}
A.~Dosovitskiy, L.~Beyer, A.~Kolesnikov, D.~Weissenborn, X.~Zhai, T.~Unterthiner, M.~Dehghani, M.~Minderer, G.~Heigold, S.~Gelly, J.~Uszkoreit, N.~Houlsby, An image is worth 16x16 words: Transformers for image recognition at scale, 2021.

\bibitem{payer2019integrating}
C.~Payer, D.~{\v{S}}tern, H.~Bischof, M.~Urschler, Integrating spatial configuration into heatmap regression based cnns for landmark localization, Medical image analysis 54 (2019) 207--219.

\bibitem{Kuhn1955}
H.~W. Kuhn, The hungarian method for the assignment problem, Naval Research Logistics Quarterly 2 (1955).
\newblock \href {https://doi.org/10.1002/nav.3800020109} {\path{doi:10.1002/nav.3800020109}}.

\bibitem{Lai2021}
K.~K.~L. Lai, T.~T.~Y. Lee, M.~K.~S. Lee, J.~C.~H. Hui, Y.~P. Zheng, Validation of scolioscan air-portable radiation-free three-dimensional ultrasound imaging assessment system for scoliosis, Sensors 21 (2021).
\newblock \href {https://doi.org/10.3390/s21082858} {\path{doi:10.3390/s21082858}}.

\bibitem{Meng2022}
N.~Meng, J.~P.~Y. Cheung, K.-Y.~K. Wong, S.~Dokos, S.~Li, R.~W. Choy, S.~To, R.~J. Li, T.~Zhang, \href{https://doi.org/10.1016/j.}{An artificial intelligence powered platform for auto-analyses of spine alignment irrespective of image quality with prospective validation}\href {https://doi.org/10.1016/j} {\path{doi:10.1016/j}}.
\newline\urlprefix\url{https://doi.org/10.1016/j.}

\bibitem{Meng2023}
N.~Meng, K.~Y.~K. Wong, M.~Zhao, J.~P. Cheung, T.~Zhang, Radiograph-comparable image synthesis for spine alignment analysis using deep learning with prospective clinical validation, eClinicalMedicine 61 (2023).
\newblock \href {https://doi.org/10.1016/j.eclinm.2023.102050} {\path{doi:10.1016/j.eclinm.2023.102050}}.

\bibitem{kunkel2011}
M.~E. Kunkel, A.~Herkommer, M.~Reinehr, T.~M. B{\"o}ckers, H.-J. Wilke, Morphometric analysis of the relationships between intervertebral disc and vertebral body heights: an anatomical and radiographic study of the human thoracic spine, Journal of anatomy 219~(3) (2011) 375--387.

\bibitem{Newell2016}
A.~Newell, K.~Yang, J.~Deng, Stacked hourglass networks for human pose estimation, Vol. 9912 LNCS, 2016.

\bibitem{Cheng2020}
B.~Cheng, B.~Xiao, J.~Wang, H.~Shi, T.~S. Huang, L.~Zhang, Higherhrnet: Scale-aware representation learning for bottom-up human pose estimation, 2020.
\newblock \href {https://doi.org/10.1109/CVPR42600.2020.00543} {\path{doi:10.1109/CVPR42600.2020.00543}}.

\bibitem{ungi2020automatic}
T.~Ungi, H.~Greer, K.~R. Sunderland, V.~Wu, Z.~M. Baum, C.~Schlenger, M.~Oetgen, K.~Cleary, S.~R. Aylward, G.~Fichtinger, Automatic spine ultrasound segmentation for scoliosis visualization and measurement, IEEE Transactions on Biomedical Engineering 67~(11) (2020) 3234--3241.

\bibitem{brignol2020automatic}
A.~Brignol, H.-E. Gueziri, F.~Cheriet, D.~L. Collins, C.~Laporte, Automatic extraction of vertebral landmarks from ultrasound images: A pilot study, Computers in Biology and Medicine 122 (2020) 103838.

\bibitem{chen2024development}
H.~Chen, L.~Qian, Y.~Gao, J.~Zhao, Y.~Tang, J.~Li, L.~H. Le, E.~Lou, R.~Zheng, Development of automatic assessment framework for spine deformity using freehand 3d ultrasound imaging system, IEEE Transactions on Ultrasonics, Ferroelectrics, and Frequency Control (2024).

\bibitem{wong2022convolutional}
J.~Wong, M.~Reformat, E.~Parent, E.~Lou, Convolutional neural network to segment laminae on 3d ultrasound spinal images to assist cobb angle measurement, Annals of Biomedical Engineering 50~(4) (2022) 401--412.

\bibitem{zhang2021flexible}
J.~Zhang, Y.~Wang, T.~Liu, K.~Yang, H.~Jin, A flexible ultrasound scanning system for minimally invasive spinal surgery navigation, IEEE Transactions on Medical Robotics and Bionics 3~(2) (2021) 426--435.

\bibitem{gueziri2020state}
H.-E. Gueziri, C.~Santaguida, D.~L. Collins, The state-of-the-art in ultrasound-guided spine interventions, Medical Image Analysis 65 (2020) 101769.

\end{thebibliography}
	%
	%
	%
\end{document}